# Magnetization Plateaus of a Double Fullerene Core/Shell Like-Nanostructure in an External Magnetic Field: Monte Carlo Study

H. Eraki, Z. Fadil, N. Maaouni, A. Mhirech, B. Kabouchi, L. Bahmad* and W. Ousi Benomar

Laboratoire de Matière Condensée et Sciences Interdisciplinaires (LaMCScI), Faculty of Sciences. P.O. Box 1014, Mohammed V University of Rabat, Morocco

**Abstract**

This paper concerns the investigation of the critical ($H_C$) and the saturation ($H_S$) magnetic fields behavior of the studied system as a function of different physical parameters. The Monte Carlo method has been used to study the magnetic properties of a ferrimagnetic behavior of a double fullerene $X_{60}$ core/shell like-nanostructure, where the symbol X can be assigned to any magnetic atom. Based on the Ising model, we focus our study on a system formed by a double sphere core/shell. The two spheres are containing the spins: $\sigma=\pm 1/2$ in the core are surrounded by the spin $S=\pm 1, 0$ in the shell. Many types of magnetization curves have been found, depending on the competitions among the exchange couplings, the crystal fields and the temperature.

______________________
* Corresponding author.
E-mail address: bahmad@fsr.ac.ma (L. Bahmad).

## 1. Introduction

A fullerene is a carbon molecule in the shape of a hollow sphere, or a line, and several other shapes. The Spherical fullerenes, also known as Buckminster fullerene or buckyballs. Cylindrical fullerenes are also referred to as carbon nanotubes (Buckytubes) [1-6]. Fullerenes are comparable in structure to Graphite, which is made of arranged Graphene sheets of connected hexagonal rings [7-10]. Twelve pentagons are essential to shut the cage and so $C_{60}$ is the only structure that best meets these experimental rules. Therefore, the $C_{70}$ presents a great solidness and stability, its structure is shaped by two parts of $C_{60}$ connected by a ring of carbon particles and it is like a rugby ball formed. Besides, there is an existence of an interminable number of conceivable varieties, from the best to the foremost outlandish and make the fullerenes an outstanding family rich of different systems [11-19]. The fullerenes have been discovered for the first time in 1985. Their discovery has led to an entirely new understanding of the behavior of sheet materials, and it has opened an entirely new chapter of Nano science and nanotechnology [20-22]. The properties of fullerenes have been examined for potential much utilize in medication, as light-activated antimicrobial specialists, also for their temperature resistance, for superconductivity and their biocompatibility. Nano materials based on fullerene have been utilized in gadgets designed for photovoltaic cells, and biomedical devices. It promptly acknowledges that these Nano systems give and accept electrons, this behavior recommends conceivable applications in batteries and progressed electronic implements [23-30]. The Ising model is one of the basic and essential models of statistical physics and material science. This model has been effectively amplified to depict different attractive Nano frameworks and plays a noteworthy part in understanding attractive magnetic properties [31-33]. There are such structures that can be or maybe well portrayed by many models [34-38], such as the Ising model which is considered in this case logical and sensible. In this research, the MCS is applied to simulate the magnetization plateaus of the fullerene type core shell structure. On the other hand, several applications including computer simulations have been developed in this field of research. Theoretically, diverse studies have been concentrated on these like nanostructures, to display their magnetic, dielectric and thermodynamic properties, via different methods with a different model of simulations [39-54]. In our several previous works, we have applied the MC simulations to examine not only the ground phase diagrams of several structures, but also the magnetic and thermodynamic properties of the different structures including the fullerene structure. This has been done under the influence of several physical parameters [55-64].

Moreover, in this paper, we investigate using Monte Carlo Simulations under metropolis algorithm, the magnetization plateaus as a function of different and important physical parameters for the double fullerene involving core shell system. This paper is following these instructions: the model and method used are illustrated in section 2, the Monte Carlo simulation details in section 3. Finally, we complete the study by a conclusion in section 4.

## 2. Model and simulations technique

Using Monte Carlo simulations, we study a ferrimagnetic double fullerene $X_{60}$ core/shell nanostructure, composed by a spin $\sigma=\pm 1/2$ in the core which is surrounded by the spin $S=\pm 1,0$ in the shell. The symbol X can be consigned to any magnetic atom, as shown in Fig. 1.

In this figure, every spin is linked to the adjacent neighbor spins with different exchange coupling interactions. Furthermore, the interaction between two identical neighboring atoms (S-S) or ($\sigma$-$\sigma$) is considered to be ferromagnetic, while the interaction between the two atoms S and $\sigma$ is considered to be anti-ferrimagnetic.

The Ising Hamiltonian of the studied system, with different kinds of spins, including the exchange coupling interactions, the external magnetic field H and the crystal field D, is given as follows:

$$\mathcal{H} = -J_C \sum_{<i,j>} \sigma_i \sigma_j - J_S \sum_{<k,l>} S_k S_l - J_{CS} \sum_{<m,n>} \sigma_m S_n - H(\sum_i \sigma_i + \sum_j S_j) - D \sum_j S_j^2 \qquad (1)$$

With $<i, j>$, $<k, l>$ and $<m, n>$ stand for the first neighbors sites (i and j), (k and l) and (m and n), respectively. Moreover, $J_C$, $J_S$ and $J_{CS}$ and are the exchange coupling constants between two first nearest neighbor atoms with spins $\sigma$-$\sigma$ (in core), S-S (in shell) and S-$\sigma$ (between core and shell), respectively.

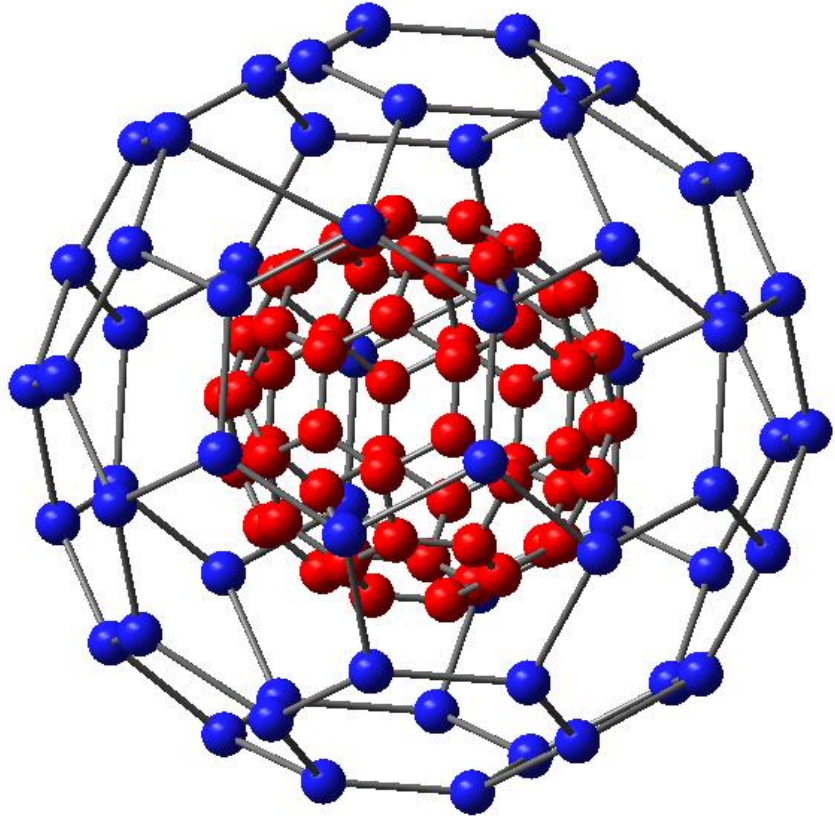

**Figure 1:** Geometry of the double fullerene core/shell like-nanostructure.

To stimulate the Hamiltonian of equation (1), we apply the standard Monte Carlo technique under the Metropolis algorithm. The studied system involves the overall number of spins $N = N_S + N_\sigma = 120$, where $N_\sigma = 60$ and $N_S = 60$ are the number of spins in the shell and the core, respectively.

To guarantee consistency, we produce the data with $10^6$ Monte Carlo steps (MCS) per site after discarding the first $10^5$ steps to reach the equilibrium, generating new configurations according to the Boltzmann distribution, in the reason to balance the present system and average over different initial conditions. Hereafter, we define the studied physical parameters as follows:

The magnetizations per spin are:

$$M_S = \frac{1}{N_S}\left\langle \sum_i S_i \right\rangle \tag{2}$$

$$M_\sigma = \frac{1}{N_\sigma}\left\langle \sum_j \sigma_j \right\rangle \tag{3}$$

The total magnetization is given by:

$$m_{tot} = \frac{M_S + M_\sigma}{2} \tag{4}$$

The internal energy per site is:

$$E = \frac{\langle \mathcal{H} \rangle}{N} \tag{5}$$

## 3. Results and discussion

In this section, we shall present some typical magnetic properties of the ferrimagnetic fullerene core/shell like-nanostructure. We focus on the effects of different physical parameters on the magnetization plateaus of this like-nanostructure, and we explore the behavior of the critical (Hc) and saturation (Hs) external magnetic fields by varying the exchange coupling interactions, the temperature and the crystal field. The typical results are presented in the figures of the section bellows.

In Figure 2a, we present the variation of the total magnetization as a function of the external magnetic field, for fixed exchange coupling coefficient values: $J_S=0.1$ and $J_C=1$, in the absence of the crystal field acting on S-spins. Thus, the following results for the magnetization plateaus were examined for T=0.3 and various exchange coupling parameter values $J_{CS} = -0.1, -0.2, -0.3$ and $-0.5$.

From this figure, all magnetizations start from the value $M_{tot}=-0.75$ and saturate for the value $M_{tot}=0.75$. Furthermore, two magnetization plateaus are localized at $M_{tot}= 0.25$ and $0.75$ corresponding to the spin configurations (-1/2, +1) and (+1/2, +1), respectively. The first plateau appears for the critical magnetic fields (Hc) while the second is reached for the saturation magnetic field (Hs). Furthermore, it is found that the critical magnetic field value Hc decreases when increasing the $J_{CS}$ parameter in its absolute values, whereas, the saturation magnetic field value Hs increases when increasing the $J_{CS}$ parameter in its absolute values.

To clarify the behavior of the (Hs) and (Hc), we collect the obtained values in Fig. 2a to plot, in figure 2b, the external magnetic fields as a function of the exchange coefficient coupling in its absolute value ($|J_{CS}|$). Indeed, when increasing the Jcs parameter in its absolute values, the magnetic field (Hs) increases while the critical magnetic field (Hc) decreases almost linearly. Such behavior is due to the anti-ferrimagnetic coefficient coupling that tends to align the spins anti-parallel of each layer existing in the core and the shell of the double fullerene.

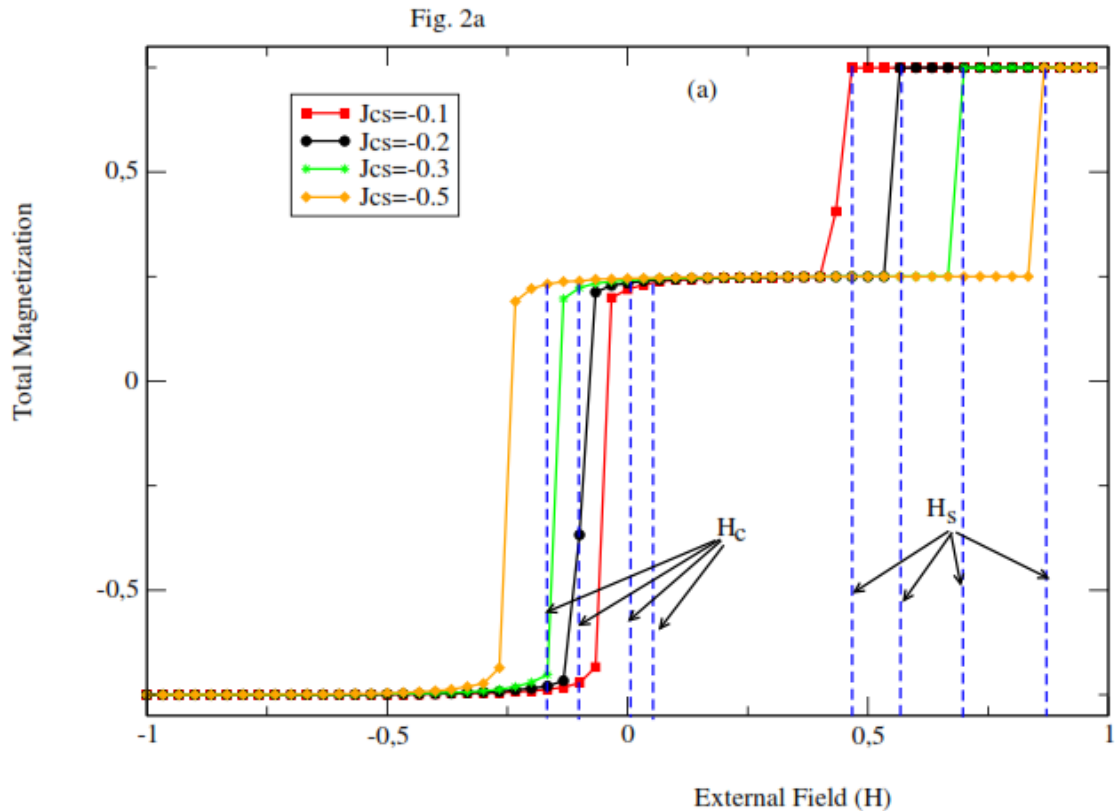

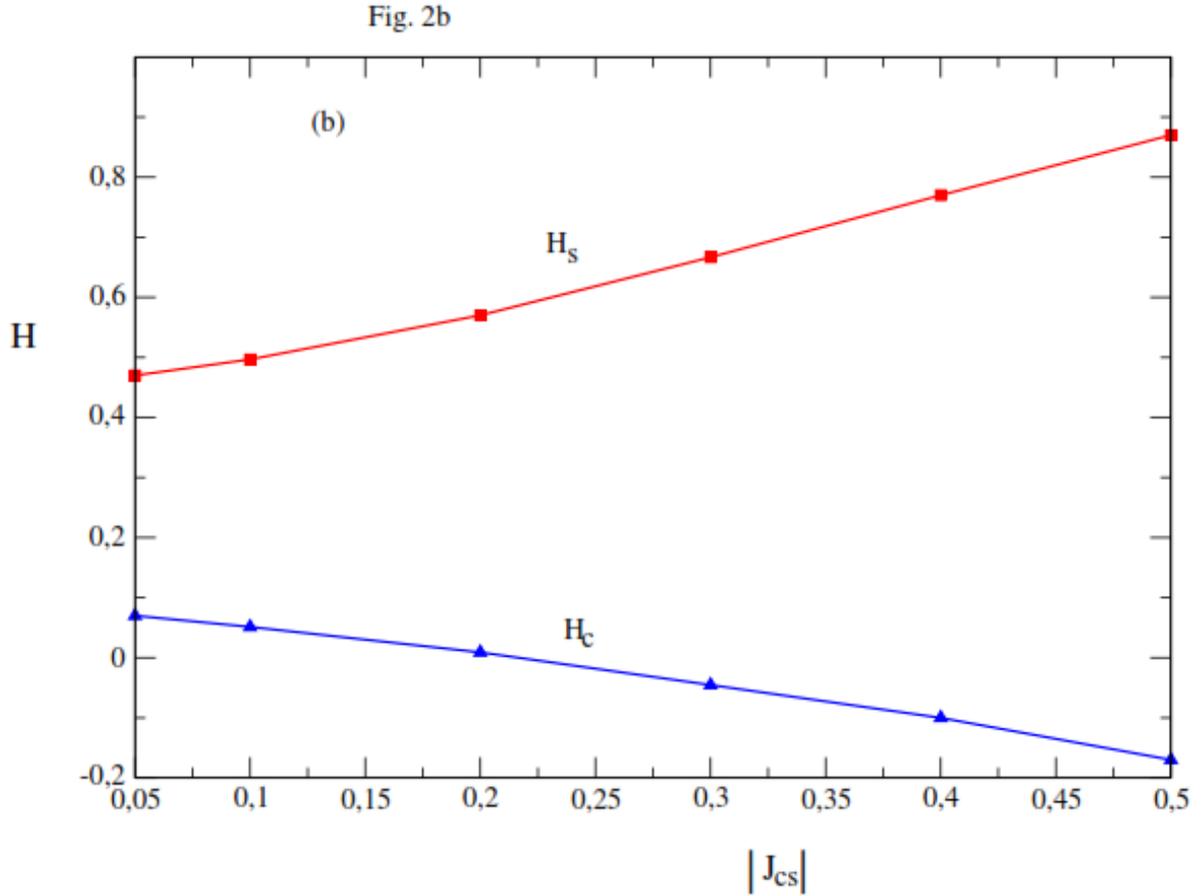

**Figure 2**: The total magnetization as a function of the external magnetic field **(a)** and the external magnetic field versus |$J_{CS}$| **(b),** for $J_C$=1, $J_S$=0.1, D=0, and T=0.3.

Figure 3a shows the total magnetization as a function of the external magnetic field for different values of the $J_S$ parameter. This figure is plotted for fixed exchange coupling coefficient values: $J_{CS}$ = -0.1, $J_C$=1, in the absence of the crystal field (D=0) and for the temperature T=0.3. We can see that all total magnetizations curves start from the same value (-0.75) corresponding to the configuration (-1/2, -1) and it remains constant, then the magnetizations undergo a first second order transition for $J_S$ < 0.1 and a first order transition for $J_S$ ≥0.1. Besides, this figure represents four magnetization plateaus (-0.25, -0.75, 0.25 and 0.75), and the distinct values of magnetic fields are highlighted by dash lines and illustrate the ($H_S$) where the total magnetization approaches the saturation value 0.75. Moreover, $H_C$ is presented by the dash lines where the total magnetization plateaus are equal to -0.25 and 0.25. Moreover, the important result obtained in Fig. 3a resides in the fact that the intermediate plateau corresponding to $M_{tot} = 0.25$ obtained for Js <0.4 is replaced by the plateau relating to $M_{tot} = -0.25$ for Js ≥0.4.

Collecting the data provided by Fig. 3a, we illustrate, in figure 3b, the behavior of the exchange coupling in the shell $J_S$ when varying the external magnetic field. We can see clearly that the critical magnetic (Hc) starts from the value 0.3, decreases and increases for Js <0.3 then increases and remains almost constant for Js ≥0.3. We notice that Hc gets remarkably influenced by the spins in the shell. Unlike (Hs) varies slightly for Js < 0.3 then increases almost linearly when Js ≥0.3. Therefore, the effect of the exchange coupling Js on the critical Hc and saturation Hs fields is significant.

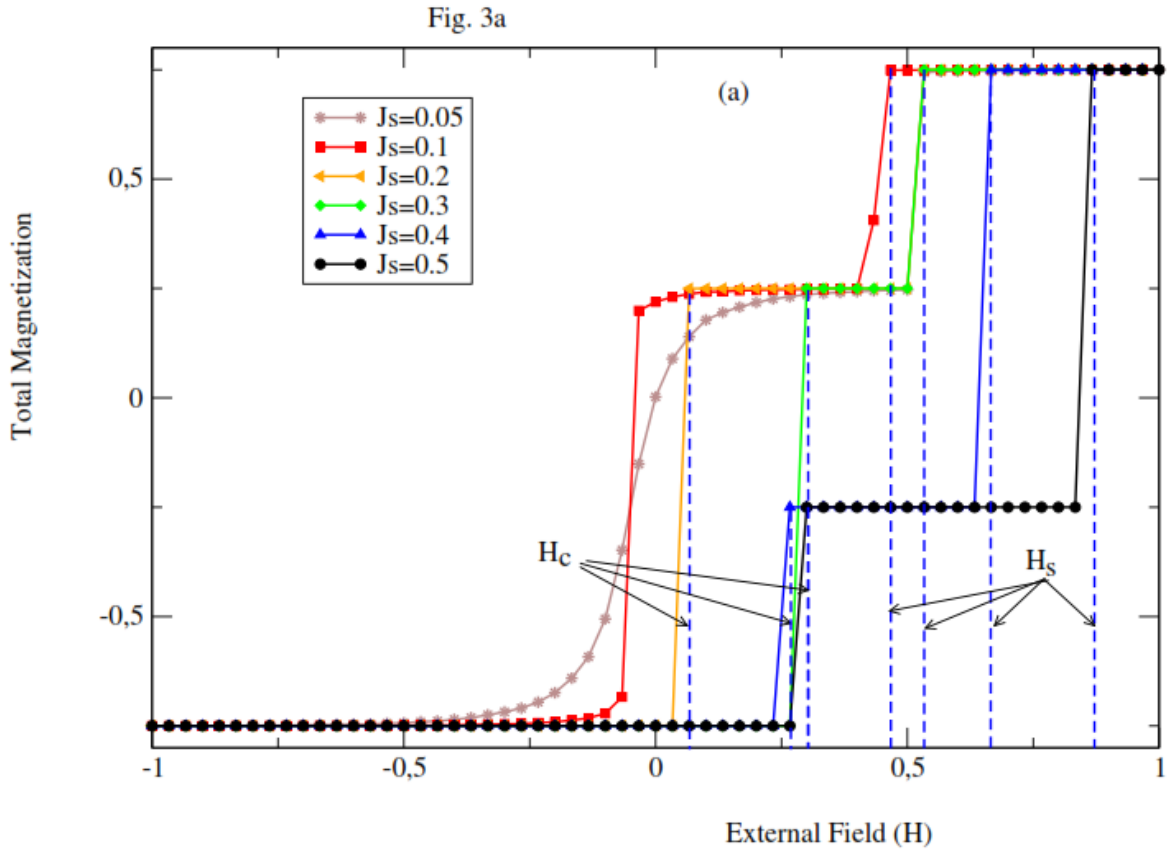

Fig. 3a

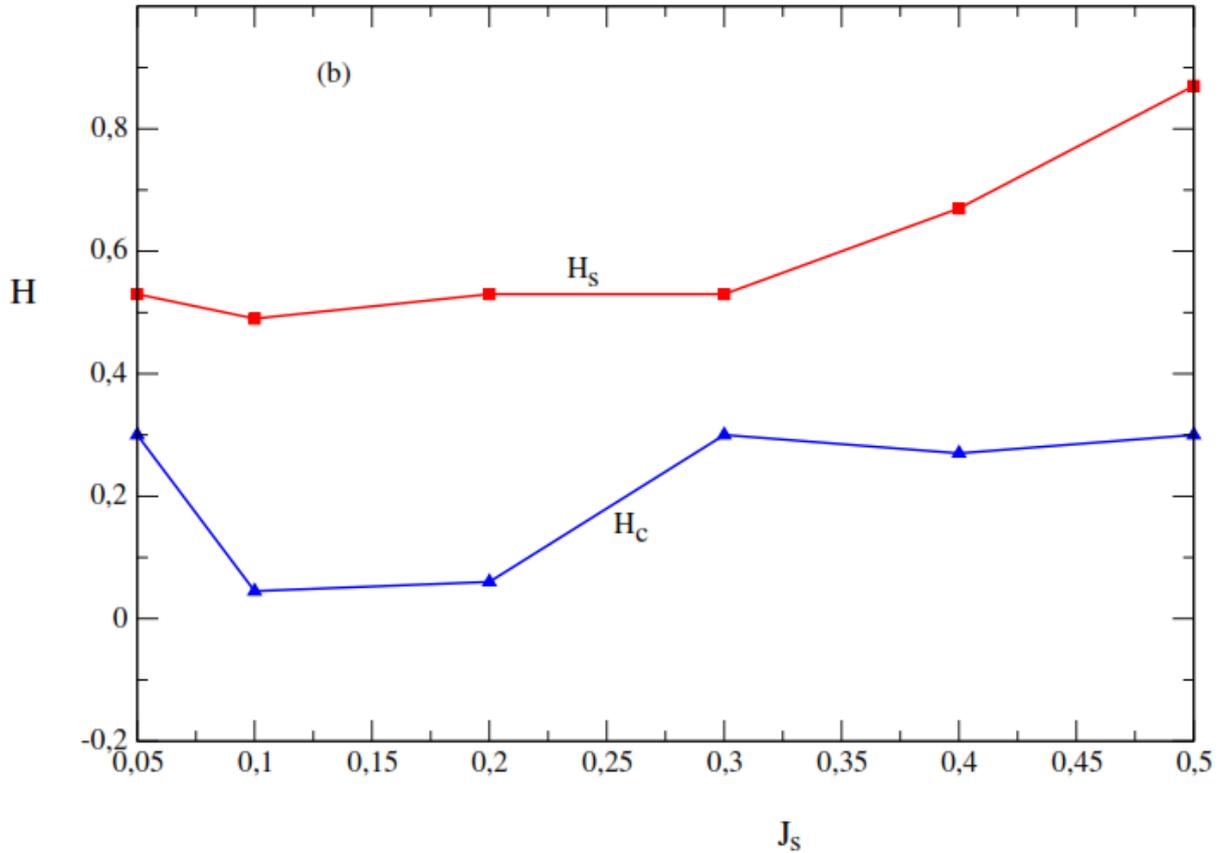

**Figure 3**: The total magnetization as a function of the external magnetic field **(a)** and the external magnetic field versus $J_S$ **(b)**, for $J_C$=1, $J_{CS}$=-0.1, D=0, T=0.3

Following the same motivation, the figures. 4a and 4b, present the effect of the exchange coupling in the core $J_C$ ($J_C$ = 0.1, 1, 2, 3, 4, and 5) on the critical Hc and saturation Hs fields for fixed parameters: $J_S$=0.1, $J_{CS}$= -0.1, D= 0, T=0.3. There is a first magnetization plateau M=+0.25 related to the configuration (-1/2, +1), corresponding to a zero critical external magnetic field value (Hc=0) independently of Jc values. Besides, when increasing the external magnetic field a second plateau is reached. However, the passage between the two plateaus takes place by a first order transition located at saturation magnetic field (Hs). Such saturation magnetic field increases when increasing the parameter $J_C$. Indeed, the saturation magnetic field values are: Hs= 0.75, 2.2, 5.1, 7.5, and 9.2 corresponding to the value of $J_C$= 1, 2 3, 4, and 5, respectively. To better illustrate this behavior, we plot figure 4b with the same values of the fixed physical parameters taken in figure 4a. It is clear that Hc remains constant even we increase the coefficient $J_C$, while the saturation magnetic field Hs increases strongly and linearly with the increase of $J_C$. Therefore, $J_C$ has no effect on the critical magnetic field Hc and the behavior of the Hs comes from the dominant effect of the external magnetic field.

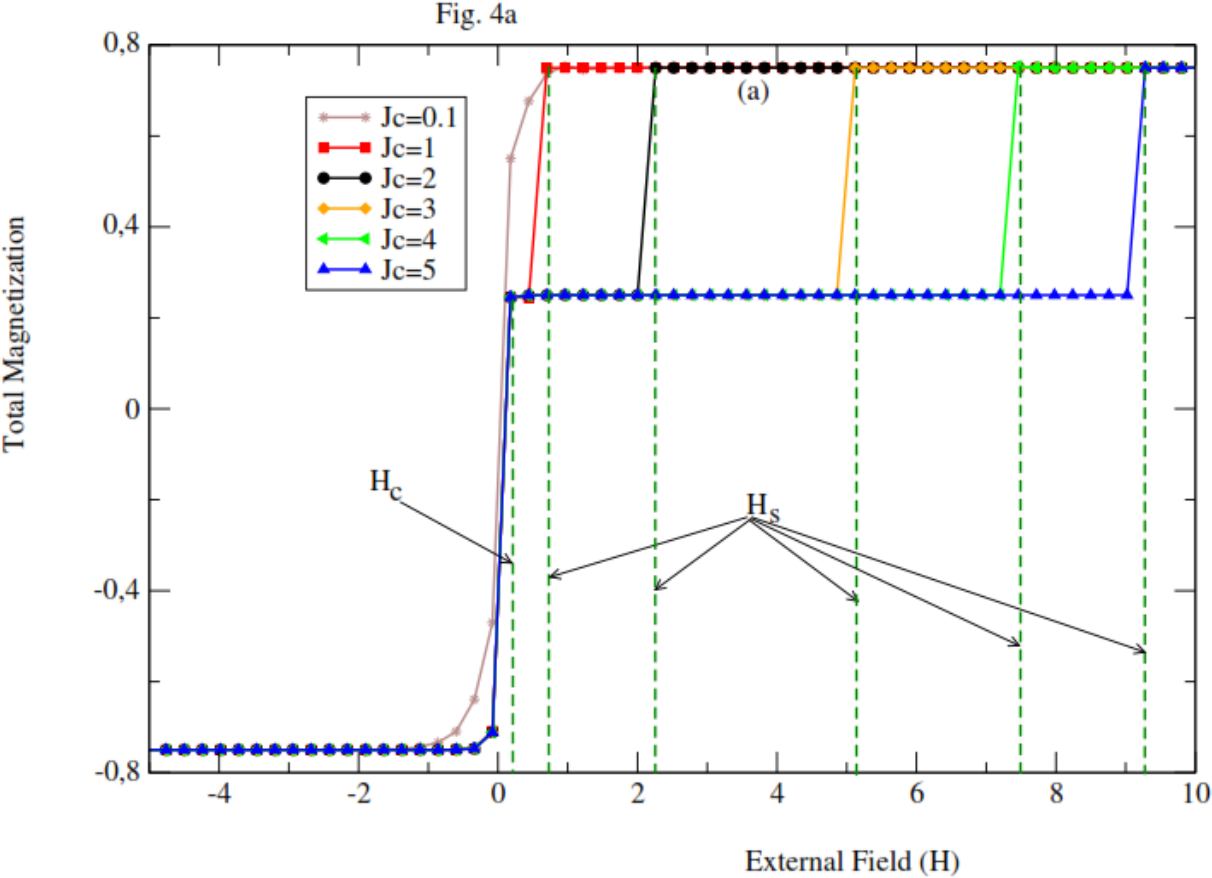

Fig. 4a

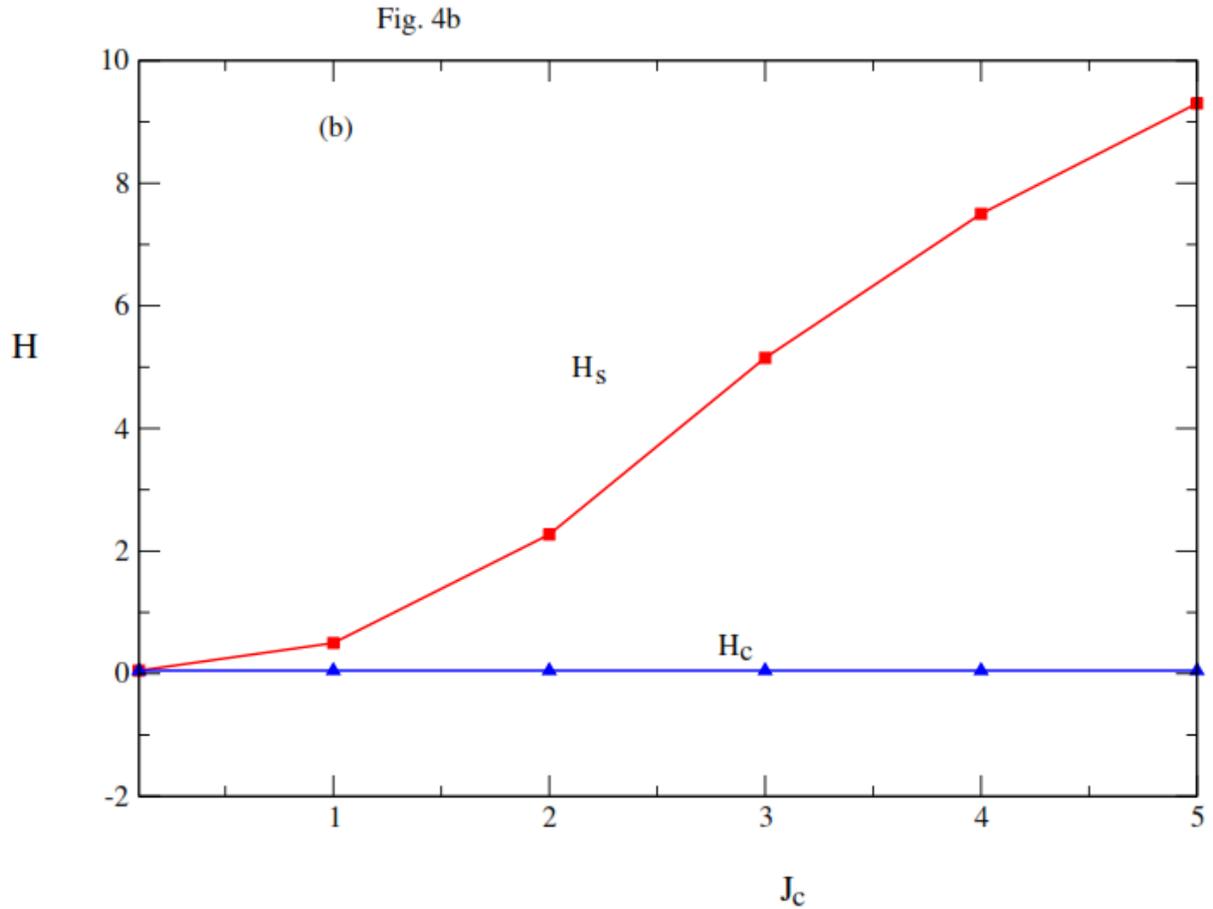

**Figure 4**: Total magnetization as a function of the external magnetic field **(a)**. The external magnetic field versus $J_C$ **(b)** for $J_S=0.1$, $J_{CS}=-0.1$, $D=0$, $T=0.3$

To focus on the effect of the temperature, we present in Figs. 5a and 5b the variation of the total magnetization versus external magnetic field for different temperature values (T=0.05, 0.1, 0.2, 0.3, 0.4 and 0.5). These figures are plotted in the absence of the crystal field and for fixed values of $J_C=1$, $J_S=0.1$, $J_{CS}=-0$. The appearance of intermediate magnetizations plateaus is found only for the configuration (-1/2, +1) corresponding to $M_{tot}= 0.25$. All values of the critical magnetic field appear for temperatures below 0.5. Indeed, for temperatures below 0.5, the critical magnetic field (Hc) varies slightly. Whereas, for temperatures greater than or equal to 0.5 the intermediate plateau disappears. On the other hand, the plateau corresponding to saturation of the total magnetization increases when the temperature is increased, inducing a decrease in HS.

Collecting the results obtained in Fig. 5a, we plot figure 5b. This figure shows the variation of Hc and Hs as a function of the temperature. Contrasting to figure 4b, in this figure Hs decreases with the temperature from Hc=2.4 until reaching the value of Hc=0.4. The results obtained from these two figures can be explained by the competition between the temperature and H since the external magnetic field tends to make the spins in a parallel direction while the temperature effect is to disorder the studied system.

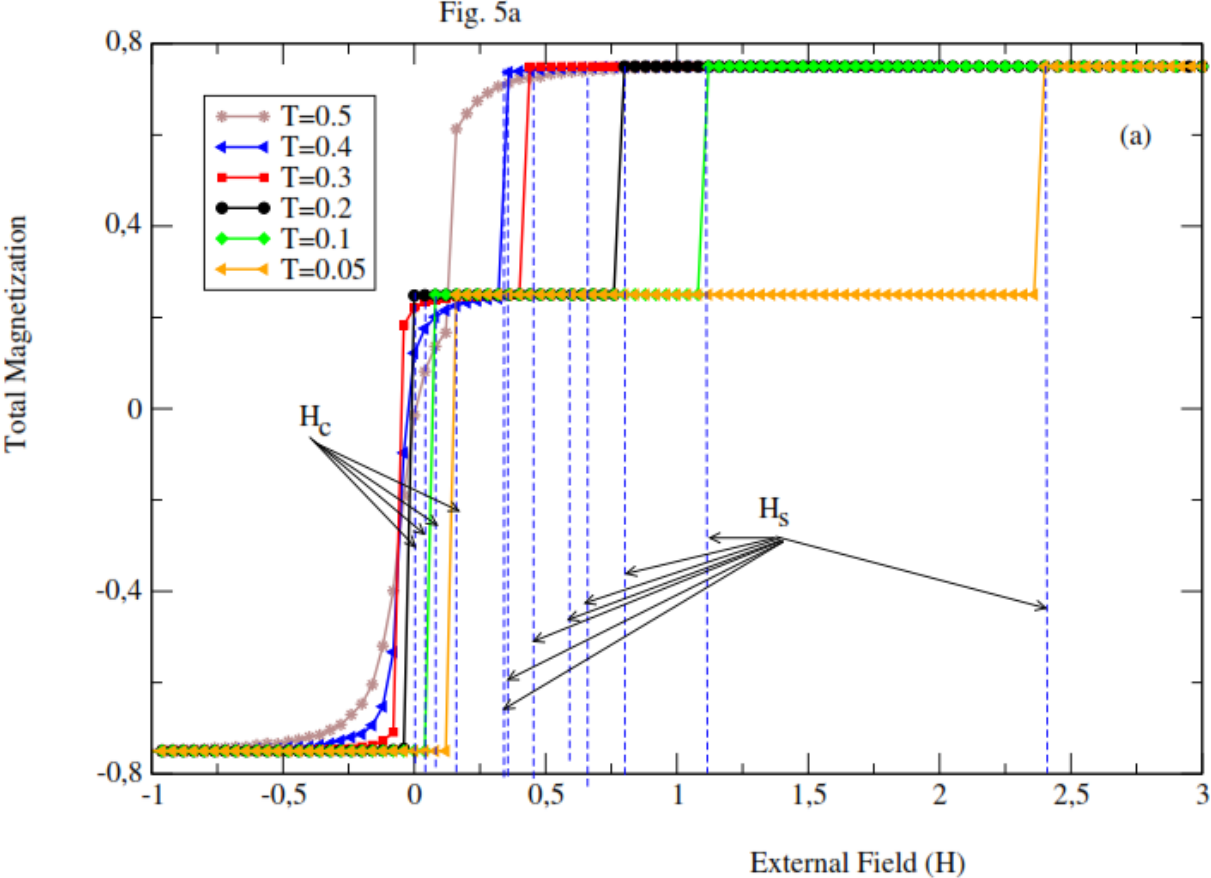

Fig. 5a

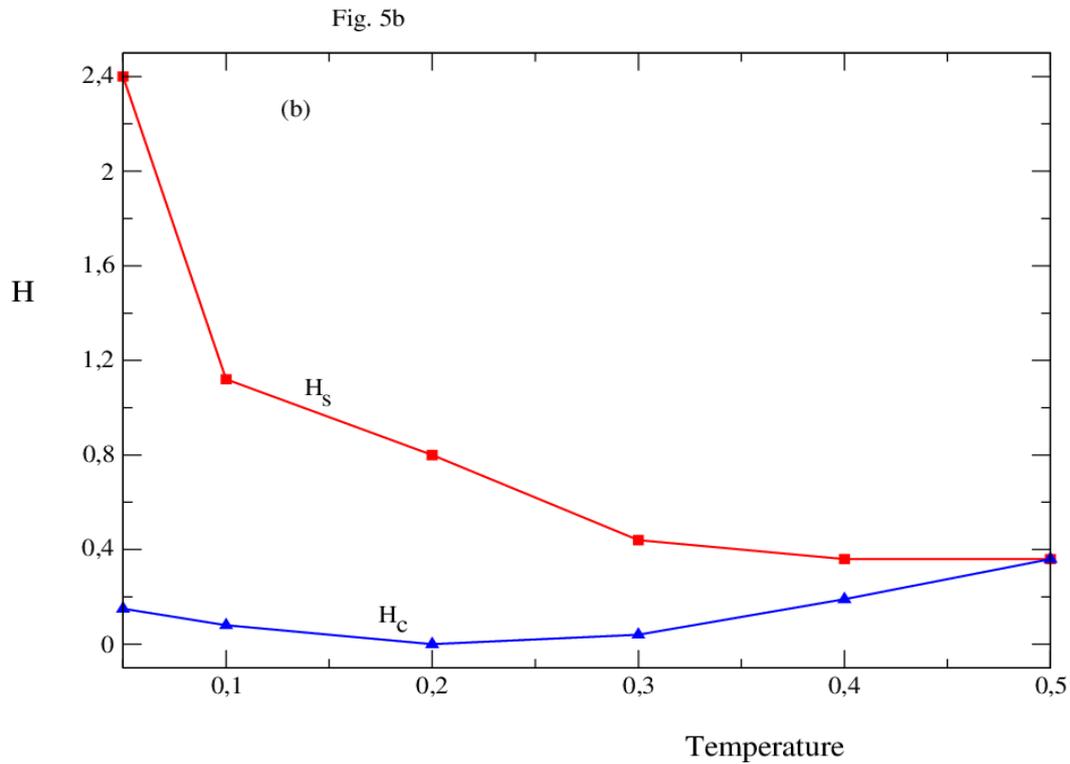

**Figure 5**: Total magnetization as a function of the external magnetic field **(a)**. The external magnetic field versus Temperature **(b)** for $J_C=1$, $J_C=0.1$, $J_{CS}=-0.1$, and $D=0$.

Finally, to complete this study, we exhibit Figs. 6a and 6b to investigate the influence of the crystal field D on the magnetization plateaus. These figures are plotted for fixed parameter values: $J_C=1$, $J_S=0.1$, $J_{CS}=-0.1$, and $T=0.3$. By varying the external magnetic field, for several values of crystal field, the total magnetization varies from -0.75 to 0.75 passing through two plateaus. The first plateau corresponds to $M_{tot} = -0.25$ and the second one to $M_{tot} = 0.25$. The transition from one plateau to another is made by first order transitions. The first plateau and the second are reached for the critical values $H_{C1}$ and $H_{C2}$ of the external magnetic field, respectively. While the last saturation plateau is reached for the value Hs. The important result revealed by Fig. 6a resides in the fact that the effect of the crystal field variation is felt only for its negative values. On the other hand, for the positive values of the crystal field, the two intermediate plateaus disappear.

The values of $H_{C1}$ increase in the interval between almost -11 to 1.8 and Hs decreases from 10.4 to 0.5, while $H_{C2}$ is almost unchangeable. These results are collected and plotted in Fig. 6b. Indeed, when D increases, $H_{C1}$ increases linearly until D= 0. While $H_{C2}$ is no affected and keeps constant, whereas $H_S$ decreases linearly all by the increase of the crystal field D until therefore these last two gathers at the point where D=0. The symmetry of the plates concerning the axis H=0 is reflected by the curves $H_S$ and $H_{c1}$. Such results can be explained by the competitions between the three physical parameters including D, H and $J_{CS}$. The external magnetic field H tends to force the spins of each layer of the core and the shell to align along their direction, whereas the antiferromagnetic exchange coupling $J_{CS}$ tends to maintain them in antiparallel alignment. On the other hand, the increasing of D favors the low states of the spins, while H favors the high spin states.

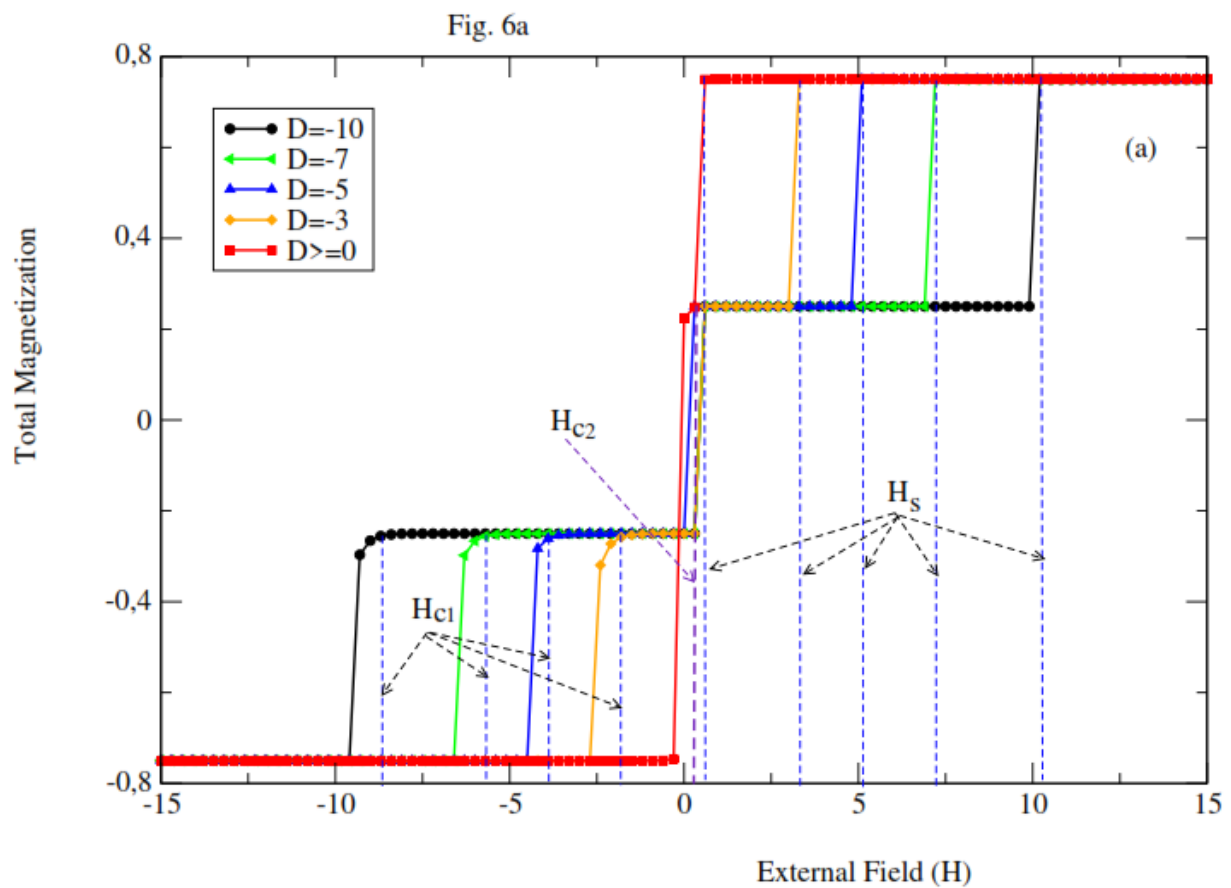

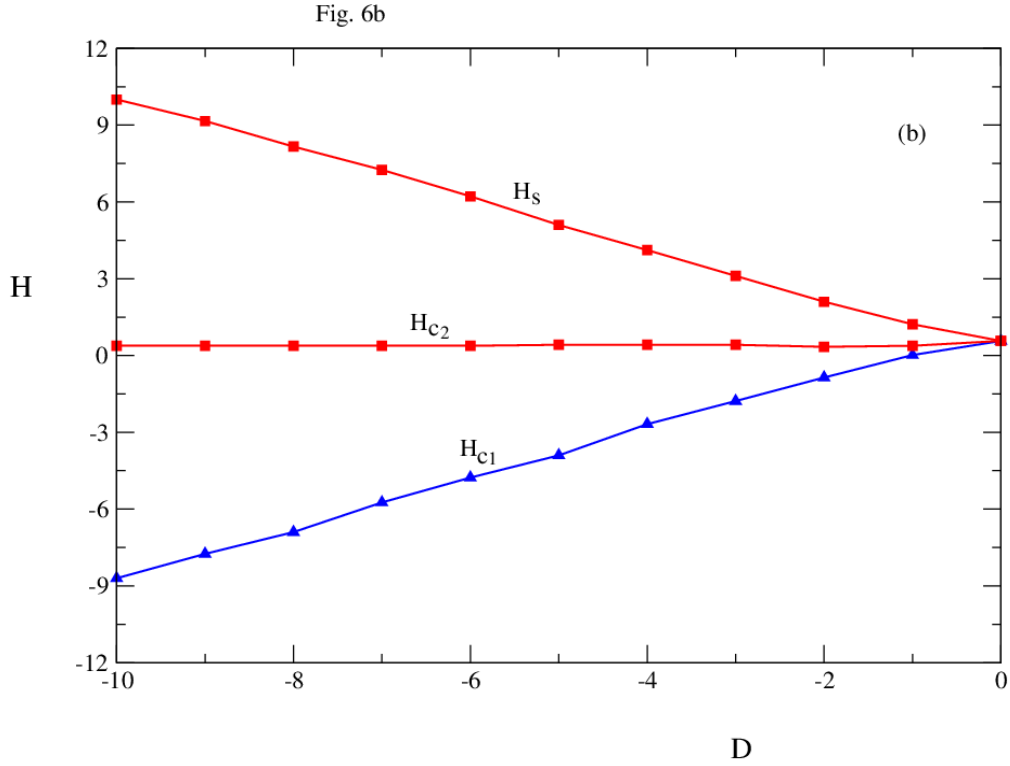

**Figure 6**: Total magnetization as a function of the external magnetic field **(a)**, the external magnetic field versus the crystal field **(b)** for: $J_C=1$, $J_S=0.1$, $J_{CS}=-0.1$, and $T=0.3$

## III. Conclusion

Using the Monte Carlo simulations, we have investigated the magnetization plateaus behaviors of a double fullerene core/shell structure under the influences of specific physical parameters: the exchange couplings, temperature and the crystal field in an external magnetic field. Additionally, the magnetization plateaus, the critical ($H_C$) and the saturation ($H_S$) magnetic fields have been investigated. It is found that the exchange coupling in the core affects strongly the saturation magnetic field, whereas the exchange coupling in the shell influences the critical magnetic field. Moreover, the crystal field has a remarkable impact on both the critical and saturation fields. Indeed, the negative values of D allow having two intermediate plateaus. These plateaus disappear when D is positive.

## References

[1] R. Abhijit. Asian Journal of Pharmaceutical Research, 2 (2012) 47.

[2] A-N. Enyashin, and A- L. Ivanovskii. Chemical Physics Letters, 473 (2009) 108.

[3] K. Iyakutti, M. Rajarajeswari, and Y. Kawazoe. Physica B: Condensed Matter, 405 (2010) 3324.

[4] C-X. Zhao, Y-Q. Yang, C-Y. Niu, J-Q. Wang, and Y. Jia. Computational Materials Science, 160 (2019) 115.

[5] Y. Lv, H. Wang, Y. Guo, B. Jiang, and Y. Cai. Computational Materials Science, 144 (2018) 170.

[6] X. Zhu, H. Yan, X. Wang, M. Zhang, and Q. Wei. Results in Physics, 15 (2019) 102738.

[7] M. Terrones, A. R. Botello-Méndez, J. Campos-Delgado, F. López-Urías & al. Nano Today, 5 (2010) 351.

[8] M. T. Lusk, and L. D. Carr. Carbon, 47 (2009)2226.

[9] S. N. Lekakh, X. Zhang, W. Tucker, H. K. Lee, T. Selly, and J. D. Schiffbauer. Materials Characterization, 158 (2019) 109991.

[10] R. Ma, Y. Zhou, H. Bi, M. Yang, J. Wang, Q. Liu, and F. Huang. Progress in Materials Science, (2020) 100665 (in proof)

[11] S. Adhikari, and R. Chowdhury. Physics Letters A, 375 (2011) 2166.

[12] M. Satoh, and I. Takayanagi. Journal of Pharmacological Sciences, 100 (2006) 513.

[13] Y. Pan, X. Liu, W. Zhang, Z. Liu, G. Zeng, B. Shao, Q. Liang, Q. He, X. Yuan, D. Huang, and M. Chen, Applied Catalysis B: Environmental, 265 (2020) 118579.

[14] T. Konno, T. Wakahara, K. Miyazawa, and K. Marumoto. New Carbon Materials, 33 (2018) 310.

[15] S. Ahmad, Chemical Physics Letters, 713 (2018) 52.

[16] J. I. Tapia, E. Larios, C. Bittencourt, M. J. Yacamán, and M. Quintana. Carbon, 99 (2016) 541.

[17] P. Bondavalli. Graphene and Related Nanomaterials: properties and applications, Elsevier, (2018) 1.

[18] S. Sachdeva, D. Singh, and S.K. Tripathi. Optical Materials, 101 (2020) 109717.

[19] P.A. Borisova, M.S. Blanter, V.V. Brazhkin, S.G. Lyapin, V.A. Somenkov, V.P. Filonenko, M.V. Trenikhin, and M.Yu. Presniakov. Diamond and Related Materials, 85 (2018) 74.

[20] R. E. Smalley. Reviews of Modern Physics, 69 (1997) 723.

[21] A.G. Avent, A.M. Benito, P.R. Birkett, A.D. Darwish, P.B. Hitchcock, H.W. Kroto, I.W. Locke, M.F. Meidine, B.F. O'Donovan, K. Prassides, R. Taylor, D.R.M. Walton, and M. van Wijnkoop. Journal of Molecular Structure, 436–437 (1997) 1.

[22] Y. Tang, J. Li, P. Du, H. Zhang, C. Zheng, H. Lin, X. Du, S. Tao. Organic Electronics, 83 (2020) 105747.

[23] H-M. Kuznietsova, N-V. Dziubenko, O-V. Lynchak, T-S. Herheliuk, D-K. Zavalny, O-V. Remeniak, Y-I. Prylutskyy, and U. Ritter. Digestive Diseases and Sciences 65 (2020) 215.

[24] Y. Yan, K. Zhang, H. Wang, W. Liu, Z. Zhang, J. Liu, and J. Shi. Colloids and Surfaces B: Biointerfaces, 186 (2020) 110700.

[25] F.G. Bernal Texca, E. Chigo-Anota, L. Tepech Carrillo, and M. Castro. Computational and Theoretical Chemistry, 1103 (2017) 1.

[26] Y. Zhang, C-R. Zhang, L-H. Yuan, M-L. Zhang, Y-H. Chen, Z-J. Liu, H-S. Chen. Materials Chemistry and Physics, 204 (2018) 95.

[27] L. Benatto, C-F-N. Marchiori, T. Talka, M. Aramini, N-A-D. Yamamoto, S. Huotari, L-S. Roman, and M. Koehler. Thin Solid Films, 697 (2020) 137827.

[28] Q-Y. Meng, B. Zhang, and D-L. Wang. Computational and Theoretical Chemistry, 1173 (2020) 112672.

[29] S. Goodarzi, T-D. Ros, J. Conde, F. Sefat, and M. Mozafari. Materials Today, 20 (2017) 460.

[30] X. Lu, M. Cui, X. Pan, P. Wang, and L. Sun. Applied Surface Science, 503 (2020) 144328.

[31] J.C. Moodie, M. Kainth., M.R. Robson and M.W. Long, Physica A: Statistical Mechanics and its Applications, 541 (2020) 123276.

[32] M. Ertaş and M. Keskin, Physica A: Statistical Mechanics and its Applications, 526 (2019) 120933.

[33] A.S. Freitas, F. Douglas, I.P. Fittipaldi and N.O. Moreno, Journal of magnetism and magnetic materials, 362 (2014) 226.

[34] Z. Berkai, M. Daoudi, N. Mendil, and A. Belghachi. Physics Letters A, 383 (2019) 2090.

[35] N. Maaouni, M. Qajjour, Z. Fadil, A. Mhirech, B. Kabouchi, L. Bahmad, W. Ousi Benomar. Physica B: Condensed Matter, 566 (2019) 63.

[36] N. Maaouni, M. Qajjour, A. Mhirech, B. Kabouchi, L. Bahmad, W. Ousi Benomar, Journal of Magnetism and Magnetic Materials, 468 (2018) 175.

[37] M. Qajjour, N. Maaouni, Z. Fadil, A. Mhirech, B. Kabouchi, W. Ousi Benomar and L. Bahmad. Chinese Journal of Physics, 63 (2020) 36.

[38] Z. Fadil, M. Qajjour, A. Mhirech, B. Kabouchi, L. Bahmad, and W. Ousi Benomar. Physica B: Condensed Matter, 564 (2019) 104.

[39] Y. Benhouria, I. Essaoudi, and A. Ainane, R. Ahuja. Physica E: Low-dimensional Systems and Nanostructures, 108 (2019) 191.

[40] Z. Fadil, N. Maaouni, A. Mhirech, B. Kabouchi, L. Bahmad and W. Ousi Benomar, International Journal of Thermophysics, 42 (2021) 1.

[41] Z. Fadil, A. Mhirech, B. Kabouchi, L. Bahmad and W. Ousi Benomar, Integrated Ferroelectrics, 213 (2021) 146.

[42] J. M. Wang, W. Jiang, C.L. Zhou, Z. Shi, and C. Wu. Superlattices and Microstructures, 102 (2017) 359.

[43] Z. Fadil, N. Maaouni, A. Mhirech, B. Kabouchi, L. Bahmad and W. Ousi Benomar, Brazilian Journal of Physics, 50 (2020) 716.

[44] Z. Fadil, A. Mhirech, B. Kabouchi, L. Bahmad and W. Ousi Benomar, Physics Letters A, 384 (2020) 126783.

[45] E-A. Moujaes, L-V. Aguiar, and M. Abou Ghantous. Journal of Magnetism and Magnetic Materials, 423 (2017) 359.

[46] Z. Fadil, A. Mhirech, B. Kabouchi, L. Bahmad and W. Ousi Benomar, Chinese Journal of Physics, 67 (2020) 123.

[47] Z. Fadil, A. Mhirech, B. Kabouchi, L. Bahmad and W. Ousi Benomar, 316-317 (2020) 113944.

[48] Z. Fadil, N. Maaouni, M. Qajjour, A. Mhirech, B. Kabouchi, L. Bahmad and W. Ousi Benomar, Phase Transitions, 93 (2020) 1.

[49] P. Jander. F.C. Santos, and S. Barreto. Journal of Magnetism and Magnetic Materials, 439 (2017) 114.

[50] C. Wu, K. L Shi, Y.Zhang and W.Jiang. Journal of magnetism and magnetic materials, 465 (2018) 114.

[51] Z. Fadil, M. Qajjour, A. Mhirech, B. Kabouchi, L. Bahmad and W. Ousi Benomar, Blume-Capel Model of a Borophene Layers Structure with RKKY Interactions: Monte Carlo Simulations, Ferroelectrics, (2021). In press.

[52] Z. Fadil, A. Mhirech, B. Kabouchi, L. Bahmad, and W. Ousi Benomar, Chinese Journal of Physics, 64 (2020) 295.

[53] Z. Fadil, M. Qajjour, A. Mhirech, B. Kabouchi, L. Bahmad, and W. Ousi Benomar, 578 (2020) 411852.

[54] Z. Fadil, A. Mhirech, B. Kabouchi, L. Bahmad, and W. Ousi Benomar, Superlattices and Microstructures, 135 (2019) 106285.

[55] S. Aouini, A. Mhirech, A. Alaoui-Ismaili, and L. Bahmad. Chinese Journal of Physics, 59 (2019) 346.

[56] Mhirech, A., Aouini, S., Alaoui-Ismaili, A., Bahmad, L., Journal of Superconductivity and Novel Magnetism, 2017, 30(11), pp. 3189–3198

[57] Masrour, R., Bahmad, L., Hamedoun, M., Benyoussef, A., Hlil, E.K., Physics Letters, Section A: General, Atomic and Solid State Physics, 2014, 378(3), pp. 276–279.

[58] Bahmad, L., Benyoussef, A., Ez-Zahraouy, H., Journal of Magnetism and Magnetic Materials, 2002, 238(1), pp. 115–122.

[59] Fadil, Z., Mhirech, A., Kabouchi, B., Bahmad, L., Ousi Benomar, W., Superlattices and Microstructures, 2019, 134, 106224.

[60] Fadil, Z., Qajjour, M., Mhirech, A., ...Bahmad, L., Ousi Benomar, W., Journal of Magnetism and Magnetic Materials, 2019, 491, 165559

[61] Mhirech, A., Aouini, S., Alaoui-Ismaili, A., Bahmad, L., Superlattices and Microstructures, 2018, 117, pp. 382–391.

[62] Belhaj, A., Jabar, A., Labrim, H., ...Laânab, L., Benyoussef, A., Solid State Communications, 2016, 226, pp. 54–59.

[63] El Yadari, M., Bahmad, L., El Kenz, A., Benyoussef, A., Journal of Alloys and Compounds, 2013, 579, pp. 86–91.

[64] Aouini, S., Mhirech, A., Alaoui-Ismaili, A., Bahmad, L., Chinese Journal of Physics, 2018, 56(4), pp. 1640–1647